\documentclass[twocolumn,showpacs,prl]{revtex4}
\usepackage{graphicx}
\usepackage{bm} % bold math
\usepackage{amssymb} % use this package to enable \nrightarrow command
\usepackage{amsmath} % use this package to enable \xrightarrow command

\begin{document}

\title{Proposed Chiral Texture of the Magnetic Moments of Unit-Cell Loop Currents in the Pseudogap Phase of Cuprate Superconductors}

\author{Sergey S. Pershoguba}
\author{Kostyantyn Kechedzhi}
\author{Victor M. Yakovenko}

\affiliation{Condensed Matter Theory Center, Department of Physics, University of Maryland, College Park, Maryland 20742-4111, USA}

\date{July 25, 2013}
%\date{v.Q, edited by SP on July 25, 2013, compiled \today}

\begin{abstract}
   We propose a novel chiral order parameter to explain the unusual polar Kerr effect in underdoped cuprates.  It is based on the loop-current model by Varma, which is characterized by the in-plane anapole moment $\bm N$ and exhibits the magnetoelectric effect.  We propose a helical structure where the vector $\bm N^{(n)}$ in the layer $n$ is twisted by the angle $\pi/2$ relative to $\bm N^{(n-1)}$, thus breaking inversion symmetry.  We show that coupling between magnetoelectric terms in the neighboring layers for this structure produces optical gyrotropy, which results in circular dichroism and the polar Kerr effect.
\end{abstract}

% Hole doped materials (cuprate superconductors), 74.72.Gh
% Kerr effect condensed matter, 78.20.Jq
% Chirality, optical activity, 78.20.Ek
\maketitle

%%%%%%%%%%%%%%%%%%%%%%%%%%%%%%%%%%%%%%%%%%%%%%%%%%%%%%%%%%%%%%%%%%%%%%%%%%%%%
\paragraph{Introduction.--} \label{sec:Intro}
%%%%%%%%%%%%%%%%%%%%%%%%%%%%%%%%%%%%%%%%%%%%%%%%%%%%%%%%%%%%%%%%%%%%%%%%%%%%%

The nature of the pseudogap phase in underdoped cuprate superconductors has been a long-standing problem \cite{Norman-2005}.  A series of optical measurements \cite{Xia-2008,Kapitulnik-2009,He-2011,Karapetyan-2012} revealed gyrotropy in this state.  It was observed that the polarizations of incident and reflected light differ by a small angle $\theta_K$, called the polar Kerr angle.  Initially, these experiments were interpreted as the evidence for spontaneous time-reversal symmetry breaking.  Theoretical models \cite{Yakovenko-2008,Kotetes-2008,Sun-2008,Moore-2012} derived optical gyrotropy from the anomalous Hall effect.  In these scenarios, the order parameter is equivalent to an intrinsic magnetic field perpendicular to the layers, which permeates the system and points inward and outward at the opposite surfaces of a crystal.  Therefore, the Kerr angle should have opposite signs at the opposite surfaces of the crystal. 

However, recent reports \cite{Armitage,Hosur-2012} found that the Kerr angle has the same sign at the opposite surfaces of a sample.  Therefore, the observed gyrotropy is not consistent with the time-reversal-symmetry breaking due to a magnetic order and should be interpreted as the evidence for natural optical activity due to chiral symmetry breaking \cite{Landau}.  Systems with helical structures, such as cholesteric liquid crystals and some organic molecules, typically exhibit optical gyrotropy and the polar Kerr effect.  It is important that the sign of the Kerr angle in this case is the same at the opposite surfaces of the system, in contrast to the gyrotropy produced by a magnetic order (see Chaps.~11 and 12 in Ref.~\cite{Landau}).

%%%%%%%%%%%%%%%%%%%%%%%%%%%%%%%%%%%%%%%%%%%%%%%%%%%%%%%%%%%%%%%%%%%%%%%%%%%%%
\begin{figure}[b]
\includegraphics[width=\linewidth]{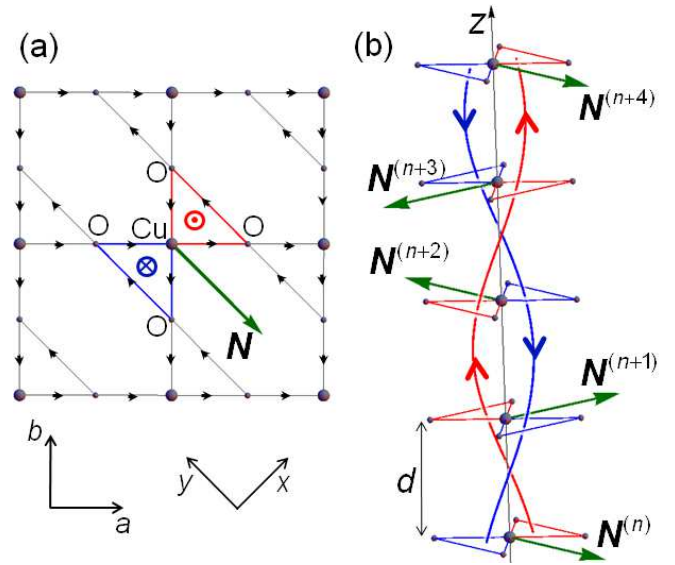}
\caption{ (a) Loop-current order in a CuO$_2$ plane \cite{Varma-2002}.  Black arrows show directions of microscopic persistent currents between copper and oxygen atoms.  Green arrow shows the anapole moment $\bm N$.
(b) Chiral order constructed on a series of parallel CuO$_2$ planes. The vector $\bm N$ rotates by the angle $\pi/2$ from one layer to another, and the period of the structure is fourfold.  The blue and red curves are the two magnetic field lines that intertwine in a double helix.} 
\label{fig:Chiral}
\end{figure}
%%%%%%%%%%%%%%%%%%%%%%%%%%%%%%%%%%%%%%%%%%%%%%%%%%%%%%%%%%%%%%%%%%%%%%%%%%%%%

Theoretical scenarios for appearance of chiral order in cuprates were proposed in Refs.~\cite{Hosur-2012,Orenstein-2012}.  Three possible orders were studied phenomenologically in Ref.~\cite{Hosur-2012}: 3-plane rotation of distorted Fermi circles, 4-plane rotation of a structure with an in-plane ferroelectric moment, and 4-plane rotation of an in-plane density wave with period 3.  A formula for gyrotropy of a chiral metal was derived in Ref.~\cite{Orenstein-2012} in terms of the Berry curvature in momentum space.  However, Ref.~\cite{Chakravarty-2013} questioned applicability of the implied assumption of coherent electron motion between the layers in Ref.~\cite{Orenstein-2012}.  X-ray circular dichroism due to chiral order was discussed in Ref.~\cite{Norman-2013}.

%%%%%%%%%%%%%%%%%%%%%%%%%%%%%%%%%%%%%%%%%%%%%%%%%%%%%%%%%%%%%%%%%%%%%%%%%%%%%
\paragraph{Helical order of the anapole moments.--}  
%%%%%%%%%%%%%%%%%%%%%%%%%%%%%%%%%%%%%%%%%%%%%%%%%%%%%%%%%%%%%%%%%%%%%%%%%%%%%

Here, we propose a novel chiral state, which could account for the polar Kerr effect in cuprates without requiring coherent electron motion between the layers.  The starting point for our construction is the model of persistent loop currents by Simon and Varma \cite{Varma-2002} shown in Fig.~\ref{fig:Chiral}(a).  The configuration of persistent currents is such that the total magnetic flux through the lattice unit cell vanishes.  The anomalous Hall effect is zero \cite{Varma-2013}, and this state does not exhibit magnetic gyrotropy.  The loop-current order \cite{Shekhter-2009} is characterized by the anapole moment $\bm N$ \cite{Zeldovich,Dubovik}, shown by the green arrow in Fig.~\ref{fig:Chiral}(a) and defined as 
\begin{equation}
	\bm N = \int d^2r\, [\bm m(\bm r)\times\bm r]
	= \frac{1}{2c} \int d^2r\, \bm r^2Ê\bm j(\bm r).
	\label{eqL}
\end{equation}
Here $\bm m(\bm r)$ and $\bm j(\bm r)$ are the microscopic densities of the magnetic moment and electric current, and the integral is taken over the unit cell.  The square symmetry of the lattice allows four possible orientations \cite{Varma-4orientations} for the vector $\bm N$, which can be obtained by progressive $\pi/2$ rotations of the configuration shown in Fig.~\ref{fig:Chiral}(a).

We propose a novel chiral state where the anapole moments $\bm N^{(n)}$ in consecutive CuO$_2$ layers labeled by the integer number $n$ are rotated by $\pi/2$, so that they trace out a helix, as shown in Fig.~\ref{fig:Chiral}(b).  This arrangement is somewhat similar to cholesteric liquid crystals \cite{director}.  This spiral structure breaks three-dimensional inversion symmetry and can be qualitatively visualized as follows.  The in-plane triangular loop currents in Fig.~\ref{fig:Chiral}(a) produce perpendicular magnetic fields of the opposite signs shown by the circled red dot and blue cross at the centers of the loops.  When the anapole moments $\bm N^{(n)}$ are arranged in the spiral structure in Fig.~\ref{fig:Chiral}(b), the red and blue magnetic field lines, propagating from one layer to another, form a double helix due to twisting of $\bm N^{(n)}$.

The chiral structure is characterized by a pseudoscalar order parameter $\Xi$ changing sign upon inversion
\begin{equation}
	\Xi =\left\langle \hat{\bm z}\cdot[\bm N^{(n)}\times\bm N^{(n+1)}]
	\right\rangle,
\label{eqAlpha}
\end{equation}
where $\hat{\bm z}$ is the unit vector along the $z$ axis.  If the anapole vector $\bm N^{(n)}$ is static, Eq.~(\ref{eqAlpha}) does not need the brackets for averaging.  However, if the vector $\bm N^{(n)}$ fluctuates, the brackets in Eq.~(\ref{eqAlpha}) represent thermodynamic and, possibly, quantum \cite{He-2012} averaging.  The chiral order parameter $\Xi$ is a local correlation function of the orientations of $\bm N^{(n)}$ in the neighboring layers and does not require long-range order in $\bm N^{(n)}$.  Spontaneous chiral symmetry breaking is known for other systems \cite{chiral-breaking,chiral-CDW}.  The configuration with $\pi/2$ rotations in Fig.~\ref{fig:Chiral}(b) maximizes $\Xi$ for a given absolute value of $N$.  Notice that Eq.~(\ref{eqAlpha}) is similar to the Dzyaloshinskii-Moriya interaction for spins and to the Lifshitz invariant $\bm N\cdot[\bm\nabla\times\bm N]$ with $\bm\nabla=\hat{\bm z}\partial_z$ in the continuous limit.

In electrodynamics of media \cite{Landau}, natural optical activity arises when inversion symmetry is broken and the expansion of the dielectric tensor $\varepsilon_{\mu\nu}(\omega,\bm k)$ in powers of the wave vector $\bm k$ has a nonvanishing first-order term
\begin{equation}
	\varepsilon_{\mu\nu}(\omega,\bm k) = \varepsilon_{\mu\nu}(\omega,0)
	+i\gamma(\omega)\epsilon_{\mu\nu z}k_z.
\label{eqEpsilon}
\end{equation}
Here $\epsilon_{\mu\nu\lambda}$ is the antisymmetric tensor, and $k_z$ is the wave vector of an electromagnetic wave propagating along the $z$ axis.  The second term in Eq.~(\ref{eqEpsilon}) represents a nonlocal effect along the $z$ axis and is responsible for gyrotropic properties of the medium.  The polar Kerr angle $\theta_K$ is determined by the following formula \cite{Bungay} to the first order in the coefficient $\gamma$ in Eq.~(\ref{eqEpsilon})
\begin{equation}
	\tan\theta_K(\omega)=\frac{\omega}{c}\,{\cal I}m
	\left[\frac{\gamma(\omega)}{1-\varepsilon(\omega)}\right].
\label{theta_K}
\end{equation}
It is clear that nonzero Kerr angle requires an imaginary part, i.e., dissipation, either in $\varepsilon(\omega)$ or $\gamma(\omega)$.  In the rest of the Letter, we derive the second term in Eq.~(\ref{eqEpsilon}) for the spiral structure in Fig.~\ref{fig:Chiral}(b).

%%%%%%%%%%%%%%%%%%%%%%%%%%%%%%%%%%%%%%%%%%%%%%%%%%%%%%%%%%%%%%%%%%%%%%%%%%%%
\paragraph*{Magnetoelectric effect in a single CuO$_2$ plane.--}
%%%%%%%%%%%%%%%%%%%%%%%%%%%%%%%%%%%%%%%%%%%%%%%%%%%%%%%%%%%%%%%%%%%%%%%%%%%%%

\begin{figure*}
\centering
\hbox{\hspace{0.8in}
\includegraphics[width=0.35\linewidth]{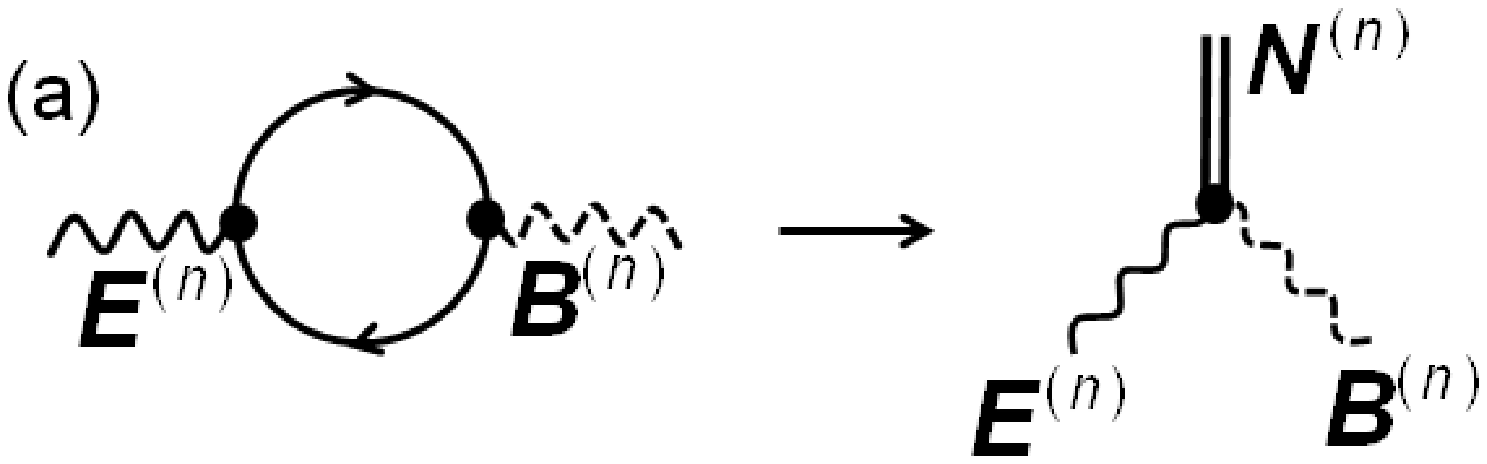} \hspace{0.5in}
\includegraphics[width=0.35\linewidth]{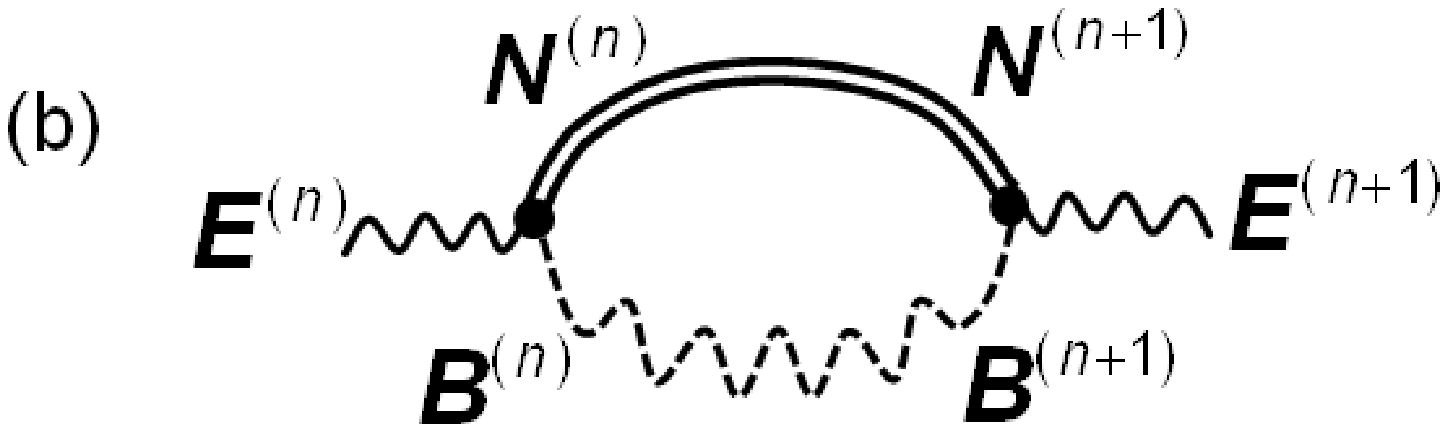} \hspace{0.5in}
}
\caption{(a) Left: Feynman diagram for the effective action of electromagnetic fields (wavy lines), obtained by integrating out the electron field (solid lines with arrows).  Right: The magnetoelectric term in the effective action, Eq.~(\ref{eqME}), where the double line represents the anapole moment $\bm N$.
(b) Coupling between the magnetoelectric terms at the neighboring layers produces the effective action for the electric field in Eq.~(\ref{NENE}).  The dashed wavy line represents the magnetic field propagator, and the double solid line represents the interlayer correlator of the anapole moments  in Eq.~(\ref{eqAlpha}).  
} 
\label{fig:Diagrams}
\end{figure*}

First let us consider a single CuO$_2$ plane with loop currents in Fig.~\ref{fig:Chiral}(a).  Integration of the electron field, schematically shown in the left diagram in Fig.~\ref{fig:Diagrams}(a), gives an effective action for the electromagnetic field with a magnetoelectric term \cite{Landau,Dzyaloshinskii}.  By symmetry, it has the form \cite{Shekhter-2009}
\begin{equation}
	S_{\rm ME}=\int d\omega\,d^2r\,\beta(\omega)\,B_z(-\omega)\,
	[\bm N\times\bm E(\omega)]_z.
\label{eqME}
\end{equation}
Here $\bm E=(E_x,E_y)$ is the in-plane electric field, $B_z$ is the out-of-plane magnetic field, and $\bm N=(N_x,N_y)$ is the in-plane anapole moment.  Given Eq.~(\ref{eqL}), the anapole moment $\bm N$ is the time-reversal-odd and parity-odd vector, so Eq.~(\ref{eqME}) has the correct symmetry structure.  It is represented graphically by the right diagram in Fig.~\ref{fig:Diagrams}(a).  Equation (\ref{eqME}) is written in the frequency representation for the electromagnetic fields, whereas $\bm N$ is taken to be static, i.e.\ having zero frequency, and $\beta(\omega)$ is a frequency-dependent coefficient.  The effective action in Eq.~(\ref{eqME}) is written in the continuous, long-wavelength limit by averaging the electromagnetic fields over distances longer than the unit cell of the lattice.

By taking a variation of Eq.~(\ref{eqME}), we find that an in-plane electric field induces an out-of-plane magnetization
\begin{equation}
	M_z(\omega) = \frac{\delta S_{\rm ME}}{\delta B_z(-\omega)} 
	= \beta(\omega)\,[\bm N\times\bm E(\omega)]_z.
	\label{eqMz}
\end{equation}
Physical interpretation is clear by symmetry in Fig.~\ref{fig:Chiral}(a). An in-plane electric field $\bm E\perp\bm N$ promotes electron transfer from one triangular loop to another, thus breaking symmetry and making one loop current stronger, which results in the net perpendicular magnetization.

Similarly, an out-of-plane magnetic field induces an in-plane electric polarization\begin{equation}
	\bm P(\omega) = \frac{\delta S_{\rm ME}}{\delta\bm E(-\omega)} 
	= \beta(\omega)\,[\bm N\times\bm B_z(\omega)].
	\label{P}
\end{equation}
The perpendicular magnetic field $B_z$ lowers the energy for one loop current and increases for another in Fig.~\ref{fig:Chiral}(a), which results in electron transfer between the loops and the in-plane electric polarization $\bm P\perp\bm N$.
%%%%%%%%%%%%%%%%%%%%%%%%%%%%%%%%%%%%%%%%%%%%%%%%%%%%%%%%%%%%%%%%%%%%%%%%%%%%%
%%%%%%%%%%%%%%%%%%%%%%%%%%%%%%%%%%%%%%%%%%%%%%%%%%%%%%%%%%%%%%%%%%%%%%%%%%%%

%%%%%%%%%%%%%%%%%%%%%%%%%%%%%%%%%%%%%%%%%%%%%%%%%%%%%%%%%%%%%%%%%%%%%%%%%%%%%
\paragraph*{Interlayer coupling and the gyrotropic term.--}
%%%%%%%%%%%%%%%%%%%%%%%%%%%%%%%%%%%%%%%%%%%%%%%%%%%%%%%%%%%%%%%%%%%%%%%%%%%%%

Now let us consider many parallel layers labeled by the integer number $n$.  In this case,  we expect a magnetic coupling between magnetizations at the neighboring layers
\begin{equation}
	S_{\rm MM} = - \int d\omega\,d^2r\, \sum_n \Lambda \, 
	M_z^{(n)}(\omega) \, M_z^{(n+1)}(-\omega).
\label{MM}
\end{equation}
This term should be derived microscopically from the distribution of electric currents inside the unit cell comparable to the interlayer distance $d$.  Here we simply write Eq.~(\ref{MM}) by symmetry for the long-wavelength fields $M_z$ with a phenomenological coefficient $\Lambda$.  We assume that the interlayer coupling  between magnetizations is instantaneous; i.e., $\Lambda$ is frequency independent. This assumption is valid when the interlayer distance $d$ is much smaller than the wavelength of light: $d\ll\lambda=2\pi/k_z$.

Substituting Eq.~(\ref{eqMz}) into Eq.~(\ref{MM}), we obtain an effective action for the electric fields in the multilayer system
\begin{align}
	S_{\rm EE} = & -\int d\omega\,d^2r\, |\beta(\omega)|^2 \sum_n \Lambda
\label{NENE} \\
    & [\bm N^{(n)}\!\times\!\bm E^{(n)}(\omega)]_z \,
	[\bm N^{(n+1)}\!\times\!\bm E^{(n+1)}(-\omega)]_z,
\nonumber
\end{align}
where we used the standard relation $\beta(-\omega)=\beta^*(\omega)$ for a linear response function.  Figure \ref{fig:Diagrams}(b) illustrates this calculation diagrammatically.  By coupling the magnetoelectric vertices shown in Fig.~\ref{fig:Diagrams}(a) and integrating out the magnetic field propagator shown by the dashed wavy line, we obtain the effective action for the electric field in Eq.~(\ref{NENE}).  The double solid line represents the interlayer correlator of the anapole moments in Eq.~(\ref{eqAlpha}).

Let us choose the $x$ and $y$ axes along the crystallographic direction $\bm a+\bm b$ and $\bm b-\bm a$ in Fig.~\ref{fig:Chiral}, so that the vectors $\bm N^{(n)}=-\bm N^{(n+2)}$ are along $x$ for odd $n$ and $y$ for even $n$.  Expanding the vector products in Eq.~(\ref{NENE}), we find two terms in the sum, for odd and even $n$.  Changing the variable $n\to n-1$ in the latter sum, we find
\begin{align}
	S_{\rm EE} = & \int d\omega\,d^2r\, |\beta(\omega)|^2 \sum_{n\;\rm odd} \Lambda \,
	N_x^{(n)} N_y^{(n+1)}
\nonumber \\
    &  E_y^{(n)}(\omega) \,
    [E_x^{(n+1)}(-\omega)-E_x^{(n-1)}(-\omega)].
\label{NNEE}
\end{align}
Using Eq.~(\ref{eqAlpha}), taking the continuous limit $z=nd$, where $E_x^{(n+1)}-E_x^{(n-1)}=2d\,(\partial E_x/\partial z)$ and $2d\sum_{n\;\rm odd}=\int dz$, and integrating by parts in $z$, we get
\begin{equation}
    S_{\rm EE} = - \frac{\Xi\Lambda}{2} \int d\omega\,d^3r\, |\beta(\omega)|^2 \, 
	\hat{\bm z}\cdot\left[\bm E(\omega)\times
	\frac{\partial\bm E(-\omega)}{\partial z}\right],
\label{eqS}
\end{equation}
or, equivalently,
\begin{equation}
    S_{\rm EE} = \frac{\Xi\Lambda}{2} \int d\omega\,d^3r\, |\beta(\omega)|^2 \, 
	\bm E(\omega)\cdot[\bm\nabla_z\times\bm E(-\omega)].
\label{rot}
\end{equation}
Comparing Eq.~(\ref{eqS}) with the standard expression
\begin{equation}
    S = \frac{1}{8\pi} \int d\omega\,d^3k\, \varepsilon_{\mu\nu}(\omega,\bm k) \,
	E_\mu(\omega,\bm k)\,E_\nu(-\omega,-\bm k),
\label{EE}
\end{equation}
we obtain the coefficient $\gamma$ in the second term in Eq.~(\ref{eqEpsilon})
\begin{equation}
	\gamma(\omega)=4\pi\Xi\Lambda |\beta(\omega)|^2.
\label{gamma}
\end{equation}
Equation (\ref{gamma}) shows that the gyrotropic coefficient $\gamma(\omega)$ is determined by the chiral order parameter $\Xi$, the interlayer magnetic coupling $\Lambda$, and the magnetoelectric coefficient $\beta(\omega)$ \cite{supplement}.  The sign of $\gamma$ depends on the sign of $\Xi$.

The above derivation was presented for equally spaced $\rm CuO_2$ layers.  However, many cuprates have the bilayer structure, where the interlayer distances alternate between $d\mp\Delta d$.  In this case, the interlayer coupling coefficient in Eq.~(\ref{MM}) is $\Lambda^{(n)}=\Lambda\pm\zeta$ for even and odd $n$.  As a result, we find an additional term, which is similar to Eq.~(\ref{NNEE}), but with $\Lambda\to-\zeta$ and $E_x^{(n+1)}\!-\!E_x^{(n-1)}\to E_x^{(n+1)}\!+\!E_x^{(n-1)}=2E_x$.  This term contributes an off-diagonal symmetric term to the dielectric tensor
$\varepsilon_{xy}=\varepsilon_{yx}=-4\pi\Xi\zeta|\beta(\omega)|^2/d$, which becomes diagonal in the crystallographic basis of $\bm a$ and $\bm b$
\begin{equation}
 \varepsilon_{aa}=-\varepsilon_{bb}=4\pi\Xi\zeta|\beta(\omega)|^2/d.
\label{nematic}
\end{equation}
Thus, we find that the helical structure in the presence of bilayers produces nematicity, i.e.\ anisotropy between the crystallographic directions $\bm a$ and $\bm b$.  This is clear by symmetry in Fig.~\ref{fig:Chiral}(b), where the pairs of layers $(n,n+1)$ have the preferred direction $\bm N^{(n)}+\bm N^{(n+1)}$ along $\bm a$.  Equations (\ref{gamma}) and (\ref{nematic}) generate circular and linear dichroism.  If $\Lambda\sim\zeta$, the linear dichroism is much stronger than the circular one because $d\ll\lambda=2\pi/k_z$.

%%%%%%%%%%%%%%%%%%%%%%%%%%%%%%%%%%%%%%%%%%%%%%%%%%%%%%%%%%%%%%%%%%%%%%%%%%%%%
\begin{figure}
\includegraphics[width=0.8\linewidth]{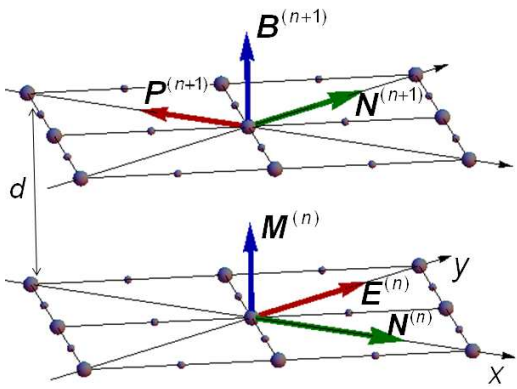}
\caption{ Schematic illustration of the rotation of electric polarization for the chiral state shown in Fig.~\ref{fig:Chiral}(b).  The electric field $\bm E^{(n)}$ at the bottom layer induces the magnetization $\bm M^{(n)}$, which couples to the top layer and induces the electric polarization $\bm P^{(n+1)}\perp\bm E^{(n)}$ because of the twist in the anapole moments $\bm N^{(n)}\perp\bm N^{(n+1)}$.} 
\label{fig:Bilayer}
\end{figure}
%%%%%%%%%%%%%%%%%%%%%%%%%%%%%%%%%%%%%%%%%%%%%%%%%%%%%%%%%%%%%%%%%%%%%%%%%%%%%

Figure~\ref{fig:Bilayer} illustrates these results graphically.  The electric field component $\bm E^{(n)}\perp\bm N^{(n)}$ generates a magnetization $M_z^{(n)}$ at the bottom layer via the magnetoelectric effect in Eq.~(\ref{eqMz}).  The magnetization $M_z^{(n)}$ produces a magnetic field $B_z^{(n+1)}$ in the same direction at the top layer via Eq.~(\ref{MM}).  This magnetic field induces an electric polarization $\bm P^{(n+1)}\perp\bm N^{(n+1)}$ at the top layer via the magnetoelectric effect in Eq.~(\ref{P}). Taking into account the third layer $n+2$ (not shown), we find that 
$\bm P^{(n+1)}\propto\hat{\bm z}\times[c_2\bm E^{(n+2)}-c_1\bm E^{(n)}]$ with some coefficients $c_1$ and $c_2$.  For equally spaced layers with $c_1=c_2$, we get Eq.~(\ref{eqS}).  For bilayers with $c_1\neq c_2$, we get the additional nematic term in Eq.~(\ref{nematic}).

%%%%%%%%%%%%%%%%%%%%%%%%%%%%%%%%%%%%%%%%%%%%%%%%%%%%%%%%%%%%%%%%%%%%%%%%%%%%%
\paragraph{Experimental relevance.--} 
%%%%%%%%%%%%%%%%%%%%%%%%%%%%%%%%%%%%%%%%%%%%%%%%%%%%%%%%%%%%%%%%%%%%%%%%%%%%%

Let us discuss possible experimental evidence for the proposed chiral order in cuprates.  Neutron scattering measurements \cite{Fauque-2006,Li-2008,Didry-2012} provide support for the loop currents shown in Fig.~\ref{fig:Chiral}(a).  However, the NMR experiments \cite{nmr-experiment} find no evidence for the local magnetic fields predicted by this model.  So, the experimental situation remains controversial.  A survey of experimental evidence supporting loop currents is presented in Ref.\ \cite{Comptes-Rendus}.

Although the loop-current order in Fig.~\ref{fig:Chiral}(a) breaks rotational symmetry in the plane as specified by the vector $\bm N$, the neutron scattering measurements \cite{Fauque-2006,Li-2008,Didry-2012} always observe the full rotational symmetry.  This may be due to domains with different $\bm N$, but the spiral order shown in Fig.~\ref{fig:Chiral}(b) also provides a natural explanation.  While the system has the tendency to break rotational symmetry in each $\rm CuO_2$ layer, we argue that it tries to restore macroscopic symmetry by orienting the vectors $\bm N$ orthogonally in the neighboring layers, which is consistent with the spiral structure.

Moreover, Refs.~\cite{Fauque-2006,Li-2008} concluded that the microscopic magnetic moments are not perpendicular to the layers, as expected from the loop currents in Fig.~\ref{fig:Chiral}(a), but have an in-plane component.  This effect can be explained by the spiral order in Fig.~\ref{fig:Chiral}(b) \cite{out-of-plane}.  Since the magnetic field lines are twisted in a double-helix structure, they are naturally tilted with an in-plane component.  In the presence of $\bm N$, the energy of the system contains the term $\bm N\cdot[\bm\nabla\times\bm B]=\bm B\cdot[\bm\nabla\times\bm N]$ \cite{Dubovik}.  Since it is linear in $\bm B$, whereas magnetic energy goes as $\bm B^2$, the system develops an equilibrium in-plane magnetic field $\bm B\propto[\bm\nabla\times\bm N]$ parallel to $\bm N$ in Fig.~\ref{fig:Chiral}(b), in qualitative agreement with Refs.~\cite{Fauque-2006,Li-2008}.  Moreover, the total energy decreases as $-|\bm\nabla\times\bm N|^2$, which favors the spiral structure.

Finally, the recent x-ray measurements \cite{Chang-2012,Ghiringhelli-2012,Achkar-2012} found doubling of the unit cell in YBa$_2$Cu$_3$O$_{7-x}$ in the $z$ direction \cite{bi-axial}.  Given the bilayer structure of YBa$_2$Cu$_3$O$_{7-x}$, the new unit cell contains four $\rm CuO_2$ layers.  The fourfold period is consistent with the spiral order shown in Fig.~\ref{fig:Chiral}(b).

For bilayer materials, the spiral structure in Fig.~\ref{fig:Chiral}(b) naturally produces nematicity, where the oxygen atoms to the left and right of the copper atom in Fig.~\ref{fig:Chiral}(a) are not equivalent to the oxygen atoms above and below.  This nematic symmetry is in qualitative agreement with the pattern observed in the scanning tunneling measurements \cite{STM-nematic}, although the same pattern was observed experimentally in bilayer and single-layer cuprates.

%%%%%%%%%%%%%%%%%%%%%%%%%%%%%%%%%%%%%%%%%%%%%%%%%%%%%%%%%%%%%%%%%%%%%%%%%%%%%
\paragraph{Conclusions.--} 
%%%%%%%%%%%%%%%%%%%%%%%%%%%%%%%%%%%%%%%%%%%%%%%%%%%%%%%%%%%%%%%%%%%%%%%%%%%%%

We propose a fourfold chiral state for cuprates obtained by twisting Varma's loop-current order by $\pi/2$ in consecutive $\rm CuO_2$ layers.  We show that this state exhibits natural optical activity and derive the gyrotropic coefficient.  It can account for the polar Kerr effect in cuprates \cite{Xia-2008,Kapitulnik-2009,He-2011,Karapetyan-2012} without invoking magnetic gyrotropy \cite{Armitage,Hosur-2012}.  For bilayer compounds, we also find nematicity and linear dichroism.  Our model is based on magnetic coupling between the $\rm CuO_2$ layers and does not require coherent electron tunneling between the layers and long-range order in the chiral structure. Other models for the polar Kerr effect in cuprates invoked magnetoelectric effects \cite{Orenstein-2011} and coupling between loop currents with different $\bm N$ \cite{Varma-2013}, but considered only a single layer, rather than the spiral multilayer structure.

%%%%%%%%%%%%%%%%%%%%%%%%%%%%%%%%%%%%%%%%%%%%%%%%%%%%%%%%%%%%%%%%%%%%%%%%%%%%%
\begin{acknowledgements}
	We thank P.~Armitage, S.~Chakravarty, S.~Davis, C.~Varma, P.~Bourges, Y.~Sidis, and M.~Greven for useful discussions.
	This work was supported by DARPA QuEST and US-ONR (K.~Kechedzhi).
\end{acknowledgements}
%%%%%%%%%%%%%%%%%%%%%%%%%%%%%%%%%%%%%%%%%%%%%%%%%%%%%%%%%%%%%%%%%%%%%%%%%%%%%

\vspace{-\baselineskip}

\onecolumngrid
\newpage % works only in the one column mode
\section*{Supplemental Material: An estimate of the Kerr angle $\bm\theta_{\rm K}$}
\twocolumngrid

There are great uncertainties in most input parameters of our model, so the very crude estimate of $\theta_K$ presented below is only an illustration, but not a definitive answer.  Equation numbers refer to the main paper.

For a crude estimate, Eq.~(4) can be written as
$$\theta_K \sim k_z\gamma = k_zd\,(\gamma/d),$$
because $\theta_K\ll1$, and the imaginary part of the dimensionless factor $[1-\varepsilon(\omega)]^{-1}$ is taken to be of the order of 1.  To estimate the dimensionless factor $k_zd$, we use the interlayer spacing in cuprates $d=1.2$~nm and the wavelength of light $\lambda=1.55$ $\mu$m in the experiments \cite{Xia-2008,Kapitulnik-2009,He-2011}:
$$k_zd=2\pi d/\lambda \approx 5\times10^{-3}.$$

For a multilayer system, Eq.~(5) should be written as a sum over layers, which becomes a three-dimensional integral in continuous limit:
$$\sum_n\int d^2r \ldots \quad \to \quad \frac{1}{d}\int d^3r \ldots $$
Then, it is convenient to introduce the magnetoelectric susceptibility in Eq.~(5)
$$\chi_{\rm ME}= \beta(\omega)N/d,$$
which is dimensionless in the Gaussian system of electromagnetic units.  Consequently, the magnetization $M_z$ in Eq.~(6) can be written as $M_z=d\chi_{\rm ME}E$.  Substituting this expression into Eq.~(8) and changing to the three-dimensional integration $(1/d)\int d^3r$, we find that $\tilde\Lambda=\Lambda d$ is the dimensionless interlayer magnetic coupling in the Gaussian system of unit.  Thus, Eq.~(14) can be written in dimensionless form
$$\gamma/d = 4\pi\, \tilde\Lambda\,\chi_{\rm ME}^2,$$
and the Kerr angle can be estimated as
$$\theta_K \sim (k_zd)\, 4\pi\,\tilde\Lambda\,\chi_{\rm ME}^2.$$

The value of $\chi_{\rm ME}$ is not known for cuprates, however we can use the following argument to estimate its magnitude.  It was much discussed in the literature \cite{Essin-2009} that a magnetoelectric coefficient in topological insulators has the value $e^2/2hc=\alpha/4\pi$, where $\alpha=e^2/\hbar c$ is the fine-structure constant.  The general consideration \cite{Essin-2010} shows that the orbital magnetoelectric coefficient for any system is typically proportional to the fine-structure constant times a dimensionless coefficient depending of the band structure.  So, we take an optimistic upper limit for the magnetoelectric susceptibility in cuprates:
$$\chi_{\rm ME} \sim \alpha=e^2/\hbar c=1/137\sim 10^{-2}.$$ \\
The coefficient $\tilde\Lambda$ is determined by magnetic interactions between microscopic orbital currents flowing within the crystal unit cell.  Because the unit cell dimensions are of the same order ($a\approx0.4$~nm and $d=1.2$~nm), and there are no other dimensionless parameters in this problem, we estimate $\tilde\Lambda$ to be of the order of 1:
$$\tilde\Lambda=\Lambda d \sim 1.$$

Collecting all results, we estimate the magnitude of the Kerr angle as
$$\theta_K \sim 2\pi\times10^{-6} \approx 6 ~\mu\rm rad.$$
The experimental value of $\theta_K$ observed in Refs.~\cite{Xia-2008-s,Kapitulnik-2009-s,He-2011-s} is a fraction of 1~$\mu$rad.  Thus, nominally, our estimate exceeds the experimental value by a factor less than 10.  However, our crude estimate can be reduced by many factors, such as the factor 
$1-\varepsilon(\omega)$ in the denominator, smaller values for the magnetoelectric coefficient $\chi_{\rm ME}$ and the interlayer coupling $\tilde\Lambda$, suppression of $\beta(\omega)$ at high frequencies, stray factors of $\pi$, etc.  Much more detailed studies and estimates of the input parameters need to be done in the future.  Nevertheless, the crude estimate presented here indicates that the proposed model may be a viable candidate for explaining the polar Kerr effect in cuprates.


\begin{thebibliography}{39}
\bibitem{director}
However, the director in cholesterics does not have an arrow, whereas the vector $\bm N$ does.  For a director without an arrow, the structure in Fig.~\ref{fig:Chiral}(b) would have two layers in the unit cell and would not be chiral.  Since the anapole moment $\bm N$ does have an arrow, the structure in Fig.~\ref{fig:Chiral}(b) has four layers in the unit cell and is chiral, because it breaks three-dimensional inversion symmetry.
\end{thebibliography}
\end{document}